\documentclass[fleqn,usenatbib]{mnras}

\usepackage{newtxtext,newtxmath}
\usepackage[T1]{fontenc}

\DeclareRobustCommand{\VAN}[3]{#2}
\let\VANthebibliography\thebibliography
\def\thebibliography{\DeclareRobustCommand{\VAN}[3]{##3}\VANthebibliography}

\usepackage{graphicx}	% Including figure files
\usepackage{amsmath}	% Advanced maths commands
\title[Longitudinal and transverse oscillations]{Stereoscopic Observation of Simultaneous Longitudinal and Transverse Oscillations in a Single Filament Driven by Two-sided-loop Jet}

\author[Tan et al.]{
Song Tan,$^{1,3}$
Yuandeng Shen,$^{1,2,3,4}$\thanks{E-mail: ydshen@ynao.ac.cn}
Xinping Zhou,$^{5}$
Zehao Tang,$^{1,3}$
Chengrui Zhou,$^{1,3}$
\newauthor
Yadan Duan$^{1,3}$
and Surui Yao$^{1,3}$
\\
$^{1}$Yunnan Observatories, Chinese Academy of Sciences, Kunming, 650216, China\\
$^{2}$State Key Laboratory of Space Weather, Chinese Academy of Sciences, Beijing 100190, China\\
$^{3}$School of Astronomy and Space Science, University of Chinese Academy of Sciences, Beijing 100049, China\\
$^{4}$Yunnan Key Laboratory of Solar Physics and Space Science, Kunming, 650216, China\\
$^{5}$College of Physics and Electronic Engineering, Sichuan Normal University, Chengdu, 610068, China
}

\date{Accepted 2013-01-18; Received 2023-01-15; in original form 2022-12-16}

\pubyear{2023}

\begin{document}
\label{firstpage}
\pagerange{\pageref{firstpage}--\pageref{lastpage}}
 \maketitle

\begin{abstract}
We report the first observations of simultaneous large-amplitude longitudinal and transverse oscillations of a quiescent filament trigged by a two-sided-loop jet formed by the magnetic reconnection between the filament and an emerging loop in the filament channel, recorded by the {\em Solar Dynamics Observatory} and the {\em Solar TErrestrial RElations Observatory}. The north arm of the jet firstly pushed the filament mass moving northwardly along the magnetic field lines consisting of the coronal cavity, then some elevated filament mass fell back and started to oscillate longitudinally at the bottom of the cavity (i.e., the magnetic dip). The northernmost part of the filament also showed transverse oscillation simultaneously. The amplitude and period of the longitudinal (transverse) oscillation are 12.96 (2.99) Mm and 1.18 (0.33) hours, respectively. By using the method of filament seismology, the radius of curvature of the magnetic dip is about 151 Mm, consistent with that obtained by the 3D reconstruction (166 Mm). Using different physical parameters of the observed longitudinal and transverse oscillations, the magnetic field strength of the filament is estimated to be about 23 and 21 Gauss, respectively. By calculating the energy of the moving filament mass, the minimum energy of the jet is estimated to be about $1.96 \times 10^{28} \operatorname{erg}$. We conclude that the newly formed jet can not only trigger simultaneous longitudinal and transverse oscillations in a single filament, but also can be used as a seismology tool for diagnosing filament information, such as the magnetic structure, magnetic field strength, and magnetic twists.
\end{abstract}


%%%%%%%%%%%%%%%%%%%%%%%%%%%%%%%%%%%%%%%%%%%%%%%%%%

%%%%%%%%%%%%%%%%% BODY OF PAPER %%%%%%%%%%%%%%%%%%

\section{Introduction}

\defcitealias{10.1093/mnrasl/slac069}{Paper I}	

Solar filaments (also known as prominences when observed at the solar limb) are relatively cold, dark materials suspended in the solar atmosphere and held together by a magnetic field \citep{2011LRSP....8....1C,2014LRSP...11....1P,2018LRSP...15....3A,2018LRSP...15....7G,2015ApJ...814L..17S}, which exhibit a variety of motions, such as oscillations \citep{2014ApJ...786..151S,2014ApJ...795..130S,2019ApJ...883..104S,2022A&A...663A..31N} or waves \citep{2007Sci...318.1577O,2018ApJ...863..192L} and even eruptions \citep{2008A&A...484..487C,2011RAA....11..594S,2012ApJ...750...12S,2021ApJ...923...45Z,2022A&A...660A.144Z}. Information on the filament magnetic field is essential to understand the formation, stability and eruption of a filament \citep{2010SSRv..151..333M}. It is generally believed that filament mass is supported in magnetic dip of sheared  magnetic arcades or magnetic flux ropes (MFRs). In some cases, oscillating filament mass can further evolve into coronal mass ejections that disturb the interplanetary space \citep{2006A&A...449L..17I,2008A&A...484..487C,2014ApJ...790..100B,2016ApJ...823L..19Z,2021ApJ...923...74D}.

\begin{figure*}
\centering
\includegraphics[width=17cm]{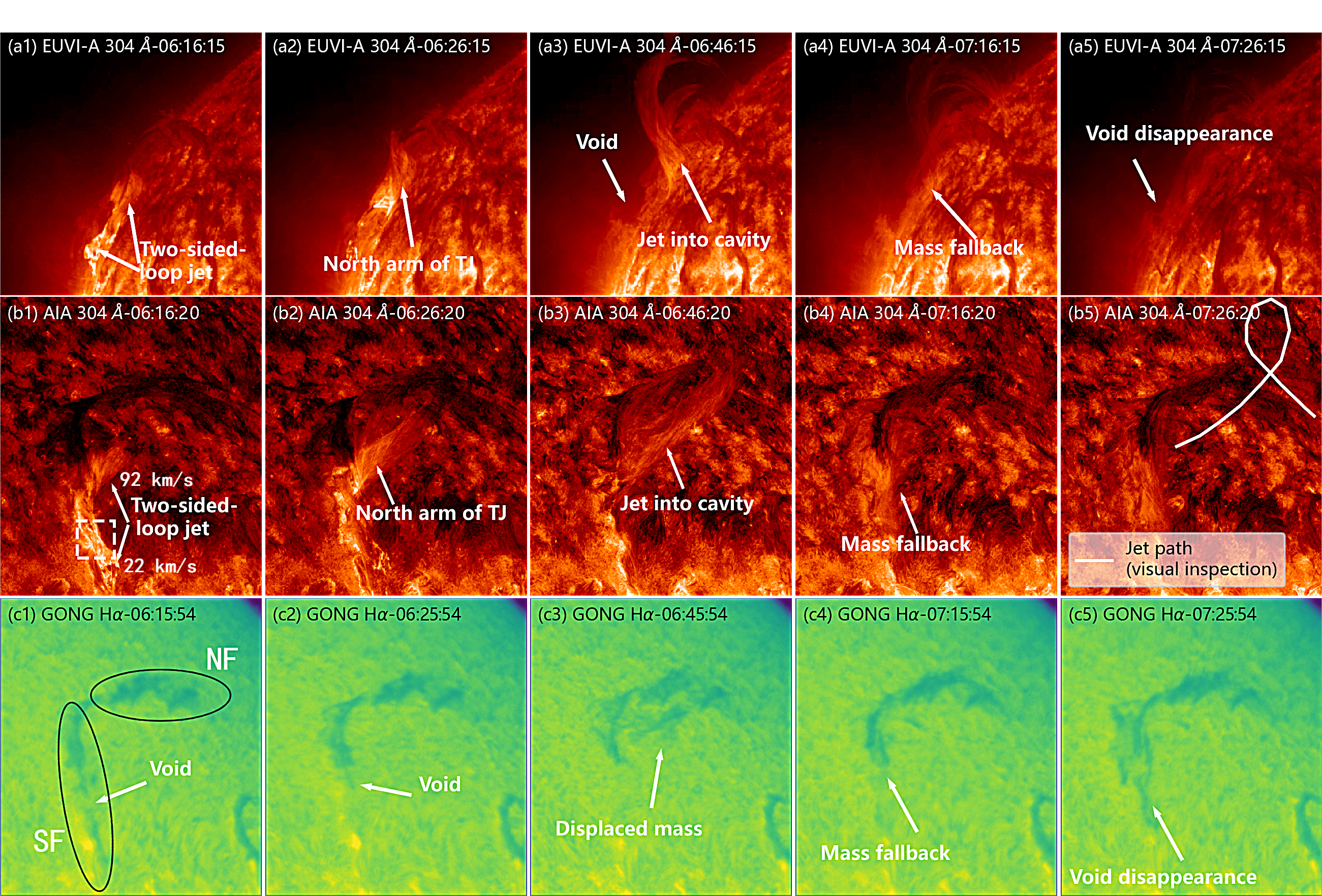}
\caption{Evolution of TJ in the filament and bifurcation of the north arm of TJ. Panels (a1)-(a5) and (b1)-(b5) are 304 \AA\ times series images from STEREO/EUVI and SDO/AIA, respectively. We have indicated the phenomena to be described with white arrows and explanatory notes. The white dashed boxes in panel(b1) indicate the source region of TJ. For the convenience of understanding and representation, we divide the filament into two parts: the southern part of the filament (SF) and the northern part of the filament (NF). The southward and northward velocities of TJ are also marked in panel (b1). The complete jet path in the cavity was also outlined with a white line in panel (b5). (An animation is provided online.)
}
\label{1}
\end{figure*}

Depending on the velocity amplitude of the oscillations (cf. reviews by \cite{2009SSRv..149..283T} and \cite{2018LRSP...15....3A}),  filament oscillations
can be classified as small amplitude oscillations (SAOs, less than 2-3$ \mathrm{~km} \mathrm{~s}^{-1}$) and large amplitude oscillations (LAOs, greater than $10\mathrm{~km} \mathrm{~s}^{-1}$). SAOs are often interpreted as propagating magnetohydrodynamic (MHD) waves, while LAOs are the periodic motions of the entire filament (or most of it) body often associated with high-energy events (e.g. flares). SAOs and LAOs have different triggering mechanisms, oscillation characteristics and physical parameters. To be more precise, we should replace "small" and "large" with "local" and "entire" to more accurately to describe filament oscillations \citep{2018LRSP...15....3A}. Here we focus on LAOs triggered by high-energy events, which can be divided into longitudinal oscillations parallel to (LALOs) and transverse oscillations perpendicular to the filament axis (LATOs, which can be further divided into horizontal and vertical oscillations relative to the solar surface), depending on the direction of the oscillations with respect to the filament axis. LALOs are often triggered by nearby flares \citep{2003ApJ...584L.103J,2007A&A...471..295V,2012ApJ...760L..10L,2012A&A...542A..52Z,2020A&A...635A.132Z}, coronal jets \citep{2014ApJ...785...79L,2017ApJ...851...47Z,2021RSPSA.47700217S}, coronal waves \citep{2014ApJ...795..130S,2017Ap&SS.362..165C}, merging of two filaments \citep{2017ApJ...850..143L}, and so on. LATOs are more intensively studied and they are often triggered by a wider variety of solar activities, such as Moreton or EIT waves \citep{2002PASJ...54..481E,2004ApJ...608.1124O,2008ApJ...685..629G,2012ApJ...745L..18A}, extreme ultraviolet waves \citep{2012ApJ...753...52L,2014ApJ...786..151S,2017ApJ...851..101S,2015ApJ...801...37T}, shock waves \citep{2014ApJ...795..130S}, coronal jets \citep{2017ApJ...851...47Z}. Please refer to the review by \cite{2018LRSP...15....3A} and statistical research by \cite{2018ApJS..236...35L}.

For LATOs, \cite{1966ZA.....63...78H} and \cite{1969SoPh....6...72K} proposed a model based on harmonic oscillators, with the restoring force supplied by the magnetic tension and with different damping mechanisms (resonant absorption, radiative damping and wave leakage). Because the model-derived period of oscillation depends on the magnetic field strength and the length of the magnetic field lines, the period of LATOs varies very widely. \cite{2003ApJ...584L.103J,2006SoPh..236...97J} reported the first case of LALOs in a filament. A complete theoretical model for LALOs was proposed by \cite{2012ApJ...750L...1L} and \cite{2012ApJ...757...98L}. For LALOs, it is generally assumed that the filament mass is stored in a magnetic dip and is perturbed to oscillate back and forth along the magnetic field direction (similar to a pendulum). The restoring force is the gravitational component of the field direction \citep{2012ApJ...757...98L,2012A&A...542A..52Z,2013A&A...554A.124Z}, and the oscillation period is determined by the radius of curvature of the magnetic dip and the cut-off period \citep{2022A&A...660A..54L}. The oscillation period of LALOs is usually around one hour, so by analyzing the oscillation parameters (e.g. the length of the period), information on the oscillation mode can also be determined. Some numerical simulation work in recent years has also contributed to our understanding of the nature of LALOs \citep{2015ApJ...799...94T,2018ApJ...856..179Z,2020A&A...633A..12M,2020A&A...637A..75L,2021A&A...654A.145L,2021ApJ...912...75L}.

In recent years, as the resolution of observations has increased significantly, simultaneous transverse and longitudinal oscillations have been widely reported \citep{2012ApJ...745L..18A,2014ApJ...795..130S,2017ApJ...851...47Z,2020A&A...633A..12M,2012ApJ...745L..18A}. The application of filament oscillation parameters to seismology analysis has also been discussed extensively \citep{2006A&A...449L..17I,2008ApJ...680.1560P,2014ApJ...786..151S,2017ApJ...850..143L}. Despite these oscillations having different triggering mechanisms  and different physical properties, they all provide powerful tools to diagnosing  the magnetic structure of filaments and coronal parameters.

\begin{figure*}
\centering
\includegraphics[width=17cm]{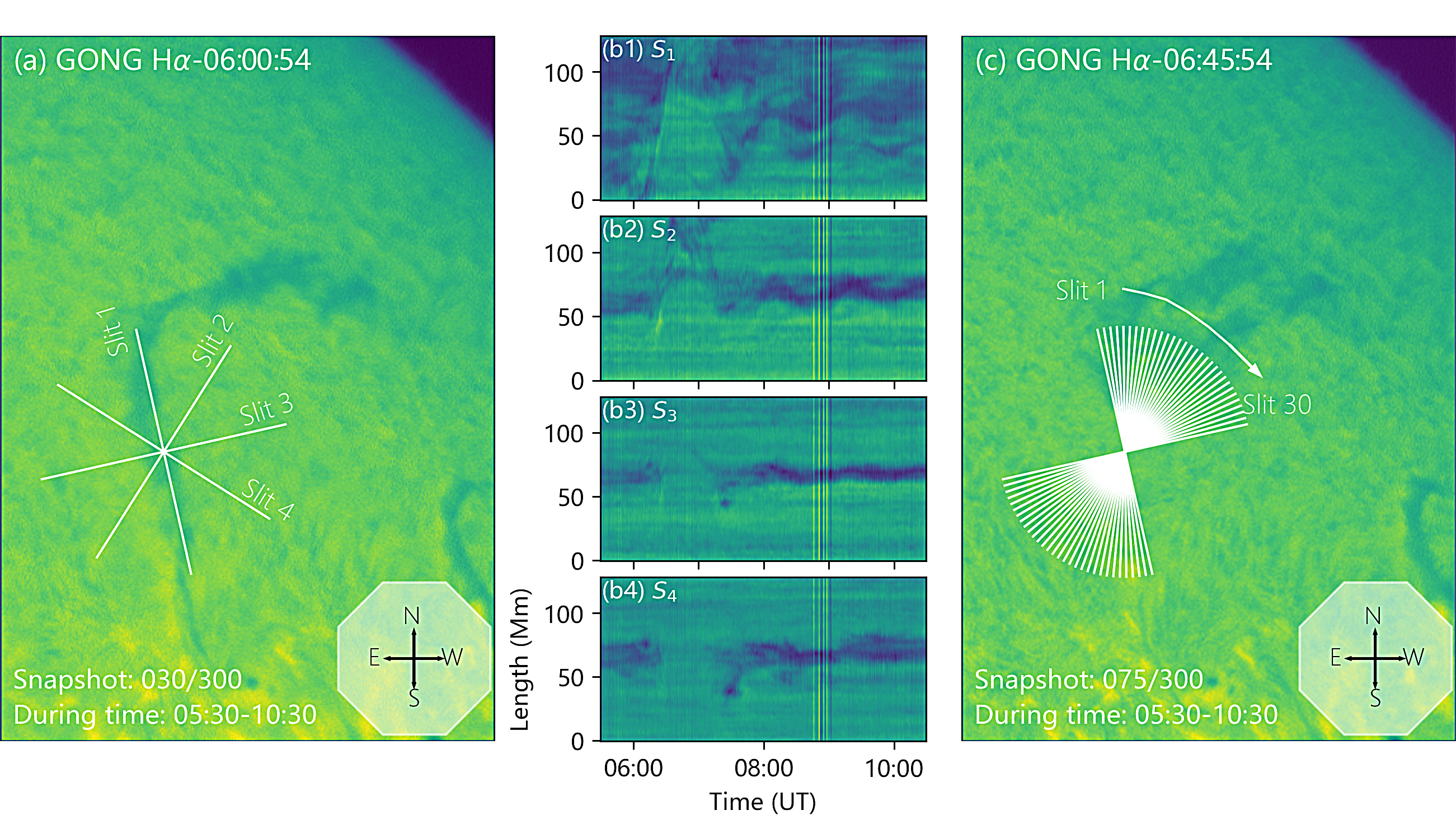}
\caption{ Panel(a) GONG h$\alpha$ image of the filament before TJ onset. We place four slits (the actual width is five pixels) at 45 degrees intervals at the centre of the filament mass motion and obtained their time-space plots (shown in panel(b)). Panel (c) GONG h$\alpha$ image of the oscillating filament. Further, to determine the actual direction of the oscillations, we place eight time-space slits between directions perpendicular to the spine and parallel to the spine. Time-space plots of thirty slits are shown in Fig. \ref{3}.
}
\label{2}
\end{figure*}

Broadly speaking, there are two types of perturbations that can trigger large-amplitude filament oscillations: the large-scale solar activity that generates shock waves \citep{2012ApJ...754....7S,2014ApJ...786..151S,2014ApJ...795..130S,2017ApJ...851..101S} and small-scale solar activity close to the filament footpoint , such as micro-flares or jets \citep{2021RSPSA.47700217S,2021ApJ...912...75L}. We are particularly interested in the latter because such small-scale solar activity is accompanied by reconnection with the magnetic structure of the filaments (e.g. reconnection-generated small jets pushing mass in magnetic dip to produce LALOs). In the past, coronal jets and filament oscillations have been studied independently. Some progress has recently been made in observations and numerical simulations for jet-triggered LALOs \citep{2014ApJ...785...79L,2017ApJ...851...47Z,2021ApJ...912...75L,2021RSPSA.47700217S}. However, the lack of stereoscopic observations leaves us with no idea of the true trajectory of the jet and how the filament oscillations are triggered. Recently \cite{10.1093/mnrasl/slac069} (hereafter called  \citetalias{10.1093/mnrasl/slac069}) stereoscopically diagnosed a two-sided-loop jet (TJ) event on 29 November 2011. They obtained the twist number of a filament by tracing the TJ pace in a filament-cavity system and validated the magnetic flux rope models (MFRs). Interestingly, the TJ also simultaneously cause the formation and recovery of filament void, suggesting the possible existence of periodic filament mass motions (i.e. filament oscillations).

In this paper, we continue to explore the filament mass motion triggered by the TJ with the aid of stereoscopic observations taken by the Atmospheric Imaging Assembly \citep[AIA;][]{2012SoPh..275...17L} onboard the Solar Dynamics Observatory (SDO) and the Extreme Ultraviolet Imager \citep[EUVI;][]{2004SPIE.5171..111W} onboard the Solar TErrestrial RElations Observatory (STEREO). The cadence and pixel size of the AIA ( 304 \AA ) images are  12 seconds and 0.6 arcsecs, respectively. The cadence and pixel size of the EUVI ( 304 \AA ) images are  10 minutes and 1.5 arcsecs, respectively. Given that filaments are more visible in the H$\alpha$ band, we also use the H$\alpha$ ( 6563 \AA ) images from the Global Oscillation Network Group \citep[GONG;][]{1996Sci...272.1284H} to analyse the mass motion. The cadence and pixel size of the GONG H$\alpha$ images are  60 seconds and 1 arcsec, respectively.

In addition, the triangulation method of SCC-MEASURE.PRO \citep{2009Icar..200..351T} allows the same features to be selected on different views of the SDO and STEREO images at the same time to generate realistic 3D coordinates of the target object. This method has also been applied in recent observations of quiescent filament \citep{2021A&A...647A.112Z,2021ApJ...911L...9G} and TJ reconstructions caused by mini-filament eruptions \citep{2019ApJ...883..104S}.

\begin{figure*}
\centering
\includegraphics[width=18cm]{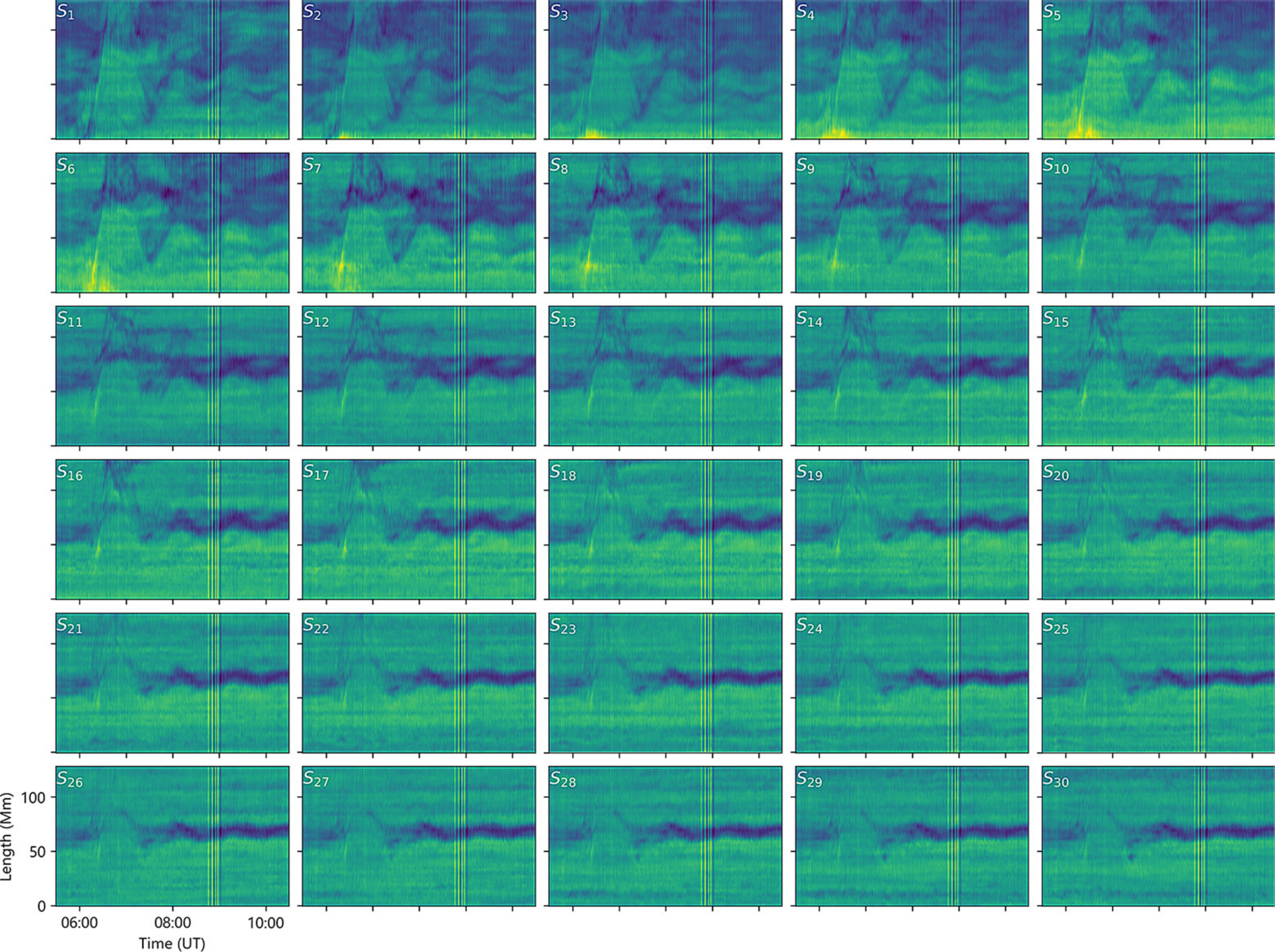}
\caption{ Thirty time-space plots of GONG H$\alpha$, spaced at an angle of 3 degrees between each and centred at the oscillating filament mass. We can see that the oscillation signal around Slit 8 are the most obvious and clear. The detailed oscillation analysis is shown in Fig. \ref{4}.}

\label{3}
\end{figure*}

\begin{figure}
\centering
\includegraphics[width=8cm]{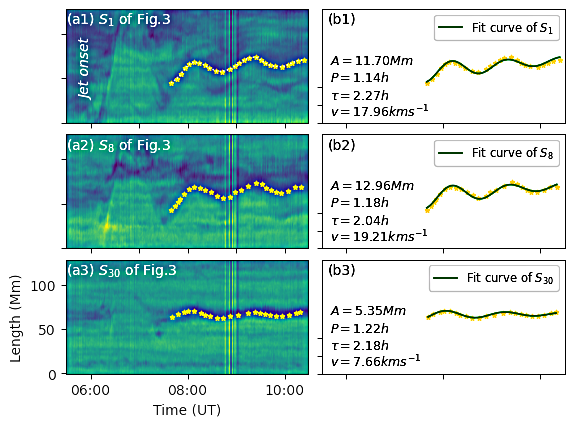}
\caption{ Panel(a) Selected time-space plots from GONG H$\alpha$ in Fig. \ref{3}. Panel(b) The position of filament centre are plotted as yellow pentagrams. The fitted functions are plotted as green lines and the useful oscillation parameters are also shown.}

\label{4}
\end{figure}

\section{Results}
\label{sec2}

\subsection{Mass motion triggered by a two-sided-loop jet}
\label{sec2.1}

TJ is bi-directional plasma flows originating from an eruption source region. There are many triggering mechanisms, including reconnection between the emergent flux loop and the horizontal field above \citep{1995Natur.375...42Y,2013ApJ...775..132J,2018ApJ...861..108Z,2019ApJ...883...52R}, reconnection of adjacent filament threads \citep{2017ApJ...845...94T}, and reconnection between the erupting mini-filament and overlying the filament system \citep{2019ApJ...883..104S,2019ApJ...887..220Y,2020MNRAS.498L.104W} or horizontal field line \citep{2019ApJ...871..220S}. \citetalias{10.1093/mnrasl/slac069} reported a TJ generated north of AR11360 that resulted from the emergence of a flux loop below a large and long (about 250 Mm) filament. The subsequent reconnection between the emerging loop and the filament resulted in the TJ (see \citetalias{10.1093/mnrasl/slac069} ).

Here, we focus on the filament mass motion (oscillations) triggered by the TJ. The whole process of jet-filament interaction is shown in Fig. \ref{1}. Refer to the online animation of Fig. \ref{1} to better understand the process. At about 06:00 UT, the two arms of the TJ (see Fig. \ref{1}(a1) and (a2)) extended from the eruption source region (white dashed boxes in Fig. \ref{1}(b1)) and moved in both north and south directions along the filament spine at velocities of $92 \mathrm{~km} \mathrm{~s}^{-1}$ and $22 \mathrm{~km} \mathrm{~s}^{-1}$, respectively. Because of the proximity to the southern footpoint, the motion of the south arm was impeded and caused  brightening near the south footpoint of the filament (see Fig. \ref{1}(c1) and (c2)). We can also use this approach to diagnose the footpoints of filament, which can be found in previous observational study \citep{2013ApJ...770L..25L}. The north arm of the TJ moved northwards along the filament spine and bifurcated at about 50 Mm from the eruption source region, with most of its mass lifting to the northwest (see Fig. \ref{1}(a2) and (b2)). The filament illuminated by the TJ allows us to obtain a twist number of about 3$\pi$ for this part of the filament (see Fig. 4 of \citetalias{10.1093/mnrasl/slac069} and its discussion). The uplifted north arm of TJ then entered the magnetic structure of the coronal cavity (see Fig. \ref{1}(a3) and (b3)) and fell along the cavity magnetic structure to the northern footpoint of the filament. The complete jet path in the cavity was outlined with a white line in Fig. \ref{1}(b5). Interestingly, there was still some of the jet mass that did not cross the apex of the coronal cavity but fell back into the magnetic dip of filament (this process is visible in the animation of Fig. \ref{1}). The motion of the jet seen in the AIA and EUVI 304 \AA\ images had a much more significant impact on the morphology of the filament in the H$\alpha$ images. During the phase of jet motion along the spine, the jet created a distinct void (see Fig. \ref{1}(c1) and (c2) and their labels) that becomes more pronounced with the motion of the north arm of TJ. As the jet entered the coronal cavity, the southern filament mass (see Fig. \ref{1}(c3)) underwent a significant displacement, even as the filament mass in the south approached the filament mass in the north. Subsequently, as the mass fell back into the filament magnetic structure, the filament morphology in H$\alpha$ was gradually recovered (see Fig. \ref{1}(c4) and (c5)). The motion of the jet and the filament mass remain synchronised throughout the process, and we believe that the TJ drove the motion of the filament mass. After the filament mass returned to its initial position, it exhibited periodic motions. In the next sections, we will focus on analysing filament oscillations.

\begin{figure*}
\centering
\includegraphics[width=18cm]{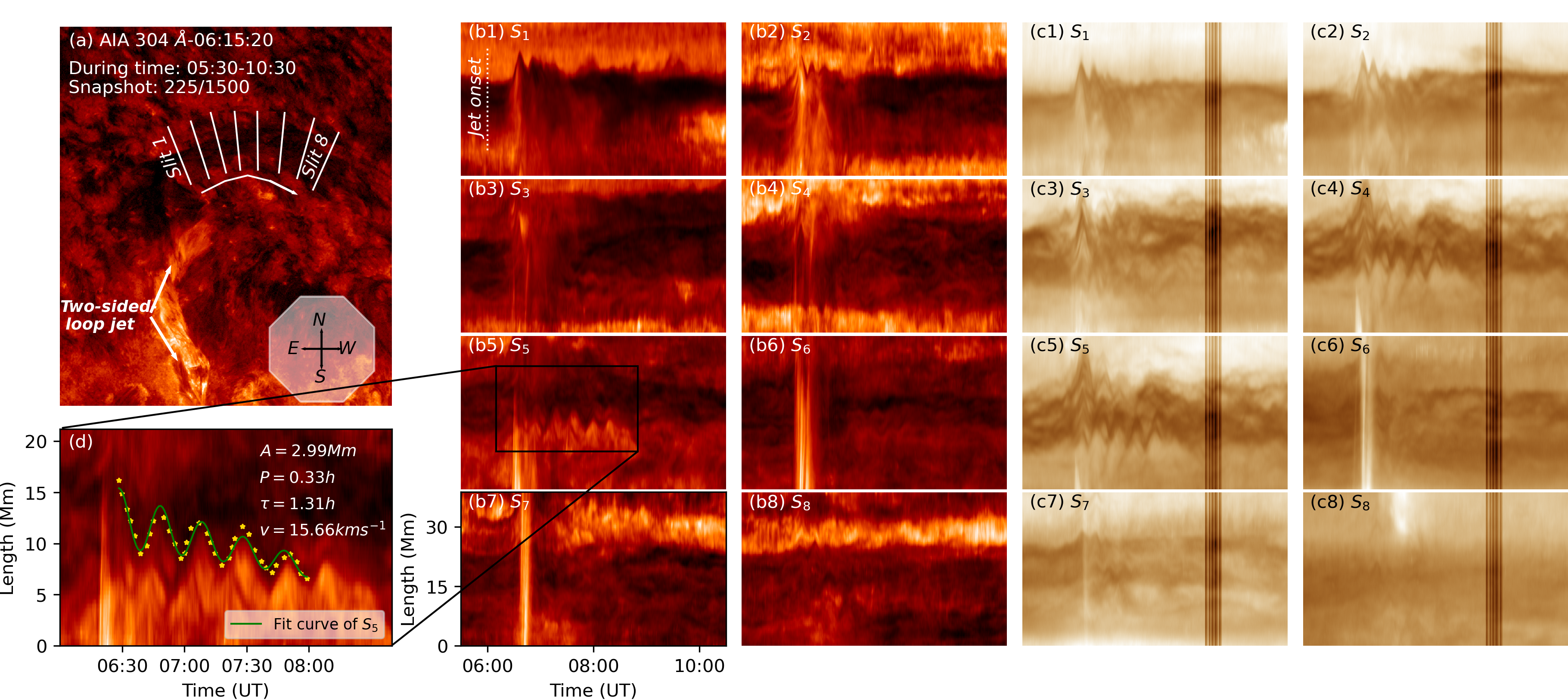}
\caption{ Panel(a) SDO/AIA 304 Å image of the filament during TJ eruption. We placed 
eight slits (the actual width is five pixels) in the northern part of the filament, perpendicular to the spine, to explore the presence of transverse oscillations. We obtained their time-space plots (shown in panel(b)). Time-space plots of 193 Å  in the same position are also shown in panel(c).
Panel (d) Selected time-space plot from SDO/AIA 304 Å images. Oscillatory signals are plotted as yellow pentagrams. The fitted functions are plotted as green lines and the useful oscillation parameters are also shown.
}
\label{5}
\end{figure*}

\subsection{Longitudinal oscillation of the southern filament}
\label{sec2.2}

The mode of filament oscillations (longitudinal or transverse) plays a crucial role in understanding the physical mechanism of the oscillations. For longitudinal oscillations, it is generally assumed that the filament mass is stored in the magnetic dip of the filament and is perturbed to oscillate back and forth along the magnetic field direction (similar to a pendulum). High-resolution observations showed that filaments are composed of long threads that are not strictly parallel to the filament spine but have an angle of 20-35 degrees (see section 2 of $Solar Prominences: Observations$ by \cite{2014LRSP...11....1P}). Based on the model of MFRs, the twisted magnetic field lines will also naturally produce an angle of divergence from the filament spine, so the direction of longitudinal filament oscillations is not exactly parallel to the spine. In previous studies, longitudinal oscillations that oscillate at an angle to the filament spine had also been mistaken for simultaneous longitudinal and transverse oscillations with the same periods \citep{2017Ap&SS.362..165C}. Therefore the mode of filament oscillation needs to be correctly determined to obtain more accurate and meaningful seismological results.

For the convenience of understanding and representation, we divide the filament into two parts: the southern part of the filament (SF) and the northern part of the filament (NF). The exact range can be found in Fig. \ref{1}(c1)). 

We firstly place four time-space slits (the actual width is five pixels, see Fig. \ref{2}(a)) across the central position of oscillating mass in the SF, with Slit 1 and Slit 3 parallel and perpendicular to the filament spine, and Slit 2 and Slit 4 at ±45 degrees (clockwise positive) to the spine, respectively. The generated time-space plots along the four slits are shown in Fig. \ref{2}(b). We can see that the Slit 1 signal of Fig. \ref{2}(b1) is more complex with oscillatory signals only partially (central and southern part of Slit 1). It is interesting to find that the oscillation phases of Slit 1, Slit 2 and Slit 3 are almost identical. To further analyse the direction of the oscillations and to obtain the oscillation parameters, we place thirty time-space slits between the parallel and perpendicular filament spine (each at an angle of about 3 degrees and centred on the oscillating mass, see Fig. \ref{2}(c)), and the results are shown in Fig. \ref{3}. Considering the resolution of the GONG H$\alpha$ image and the width of the slit, the thirty slits placed are already more than the actual resolution and sufficient for us to accurately judge the direction of the longitudinal oscillation.

We can see that the oscillation signal around Slit 8 are the most obvious and clear in Fig. \ref{3}. Moreover, the further away from Slit 8, the smaller the amplitude, suggesting that Slit 8 may represent the actual direction of mass oscillations. Therefore, we believe this is a longitudinal oscillation with an angle between the direction of the oscillation and the filament spine. To test our idea, we carefully analyse the oscillation parameters of Slit 8 with respect to the parallel/perpendicular filament spine direction.

We manually mark the positions of the filament in Fig. \ref{4}(a) with yellow pentagrams and fitted the curves with the widely used equation,
\begin{equation}
s = A \sin \left(\frac{2 \pi}{P} t+\phi\right) e^{-t / \tau}  + b t+s_{0}
\end{equation}
where $A$, $\phi$, and $s_{0}$ represent the initial (about 07:45 UT)  amplitude, position, and phase respectively. $b$, $P$, and $\tau$ indicate the linear velocity of the filaments, period, and damping timescale of the oscillations. We also calculate the velocity of the filament oscillations ($v = ds/dt$), the useful parameters of which are shown in Fig. \ref{4}(b). Here we compare the relationship between the amplitudes of Slit 8, Slit 1 and Slit 30, and we find that $\sqrt{A_{1}^{2}+A_{30}^{2}} \approx  12.86$ Mm, which is very close to $A_{8}$ (12.96 Mm). Further, Slit 1, Slit 8 and Slit 30 have very similar oscillation periods (1.14 h, 1.18 h, and 1.22 h) and damping timescales (2.27 h, 2.04 h, and 2.18 h). So we can consider that the oscillations parallel (Slit 1) and perpendicular (Slit 30) to the spine are both components of the real oscillation (Slit 8). Therefore, we conclude that this is a LALO of the SF with an angle (about 21 degrees) to the spine (without considering projection effects).

\subsection{Transverse Oscillation of the northern filament}
\label{sec2.3}
In previous high-resolution EUV observations, \cite{2017ApJ...851...47Z} found that distant jets triggered different parts of the filament to exhibit different oscillations. The eastern part of the filament they studied exhibited longitudinal oscillations, while the western part exhibited transverse oscillations. This suggested that the jet triggered  the filament's simultaneous longitudinal and transverse oscillations. Unlike previous work, the jets in our study originated from the magnetic reconnection between the emerging flux loop and filament inside the filament channel. In Section \ref{sec2.2}, we focus on the LALO of the SF; the presence of oscillations in the NF still needs to be analysed. Considering that in the H$\alpha$ images, the NF did not appear to be voided and recovered as in the SF, we suggest that it may be the transverse oscillations.

We place eight slits uniformly in the NF perpendicular to the spine in AIA 304 Å and 193 Å (see Fig. \ref{5}(a)) to analyse the transverse oscillations. The results of the eight time-space diagrams are shown in Fig. \ref{5}(b) and (c). We can see that almost all diagrams exhibit oscillatory signals and that the oscillations are generated almost simultaneously with the jet onset (see Fig. \ref{5}(b1)). In particular, for Slit 5 (both in 304 Å and 193 Å), a complete oscillation signal lasting four or five cycles is exhibited. Using the same oscillation equation in section \ref{sec2.2}, we fit the oscillation signal to obtain the oscillation parameters (see Fig. \ref{5}(c)). We find that the amplitude (2.99 Mm) and period (0.33 h) of the NF are significantly smaller than those of the SF. Moreover, the onset of the transverse oscillation in the northern part (at about 6:20 UT) was much earlier than longitudinal oscillations in the southern part (at about 07:45 UT). Thus, we conclude that the NF had a different transverse oscillation than the SF. Thus, by analysing the time-space slits in different filament parts, we find that this TJ triggers simultaneous longitudinal and transverse oscillations in a single filament.

For a jet moving along a filament magnetic structure, it is natural to trigger longitudinal oscillations. And if we consider that the jet is moving northwards at $92 \mathrm{~km} \mathrm{~s}^{-1}$ and that the length of the SF is about 100 Mm, then the jet reaches the NF at about 06:18 UT, which corresponds well with the onset time of the transverse oscillation (about 06:20 UT). The question arises then, how does the jet trigger transverse oscillations of the northern filament? We believe that first of all, the generation of the jet caused a northward displacement of the filament as a whole. The motion of the jet, which then moves partially along the filament spine, is impeded, and in the process, the kinetic energy of the jet is transformed into the kinetic energy of the filament oscillations. In this case, the NF deviating from the equilibrium position exhibited LATO in the case of magnetic tension as a restoring force. Due to the lack of high-resolution observations, the triggering detail of transverse oscillations, where the oscillation signal is relatively weak in Fig. \ref{5}, is still worth researching later.

\section{Discussions}
Having obtained the oscillation
parameters, we will then perform the filament seismology analysis and give our understanding of the physical picture of the jet-triggered oscillation.

\subsection{Filament seismology analysis}
\label{sec3.1}

\subsubsection{Radius of curvature}
\label{sec3.1.1}

Both theoretical solutions and numerical simulations demonstrated that the main restoring force of filament LALOs is the gravity component along the direction of the magnetic dip field lines \citep{2012ApJ...750L...1L,2012ApJ...757...98L,2012A&A...542A..52Z,2018ApJ...856..179Z}. According to the extended pendulum model of longitudinal oscillations \citep{2022A&A...660A..54L}, the following relationship exists between the radius of curvature and the period,
\begin{equation}
R=\frac{g_{0} P^{2}}{4 \pi^{2}\left[1-\left(\frac{P}{P_{\odot}}\right)^{2}\right]}
\end{equation}
where $g_{0}$ ($274 \mathrm{~m} \mathrm{~s}^{-2}$) is the gravitational acceleration at the solar surface, and $P_{\odot}$ is the cut-off period (2.78 h). This brings in the oscillation period (1.18 h), which gives a radius of curvature (about 151 Mm).

\begin{figure*}
\centering
\includegraphics[width=18cm]{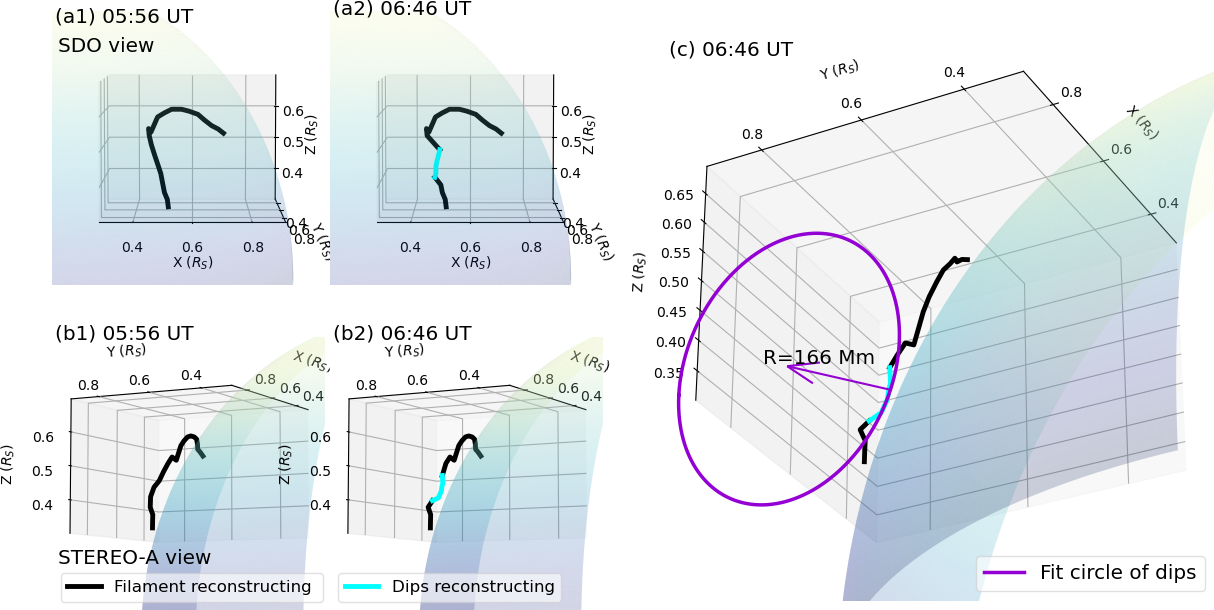}
\caption{3D filament reconstruction results. Panels (a) and (b) are 3D reconstructions of filament from SDO and STEREO views, respectively.
The black curve in panels (a1) and (b1) indicate the undisturbed filament. The cyan curve in panels (a2) and (b2) indicate the  part of the magnetic dip. (c1) We fit a circle to the bottom of the magnetic dip (represented by the cyan curve) to find its radius of curvature (about 166 Mm).
}
\label{6}
\end{figure*}

Here, with the aid of stereoscopic observations, we present a new method for estimating the radius of curvature of magnetic dip. In the results of section \ref{sec2.1}, we have described the formation of the void by the jet. From the STEREO observations, we can clearly see that the jet pushed away the filament mass, which allowed us to "see" the magnetic dip directly (see Fig. \ref{1}(a2)). We reconstruct the unperturbed and perturbed filament using stereoscopic observations taken by the SDO/AIA and the STEREO/EUVI 304 Å paired images, and the panels (a) and (b) of Fig. \ref{6} show the reconstruction results seen from the SDO and STEREO views, respectively.

\begin{figure}
\centering
\includegraphics[width=8cm]{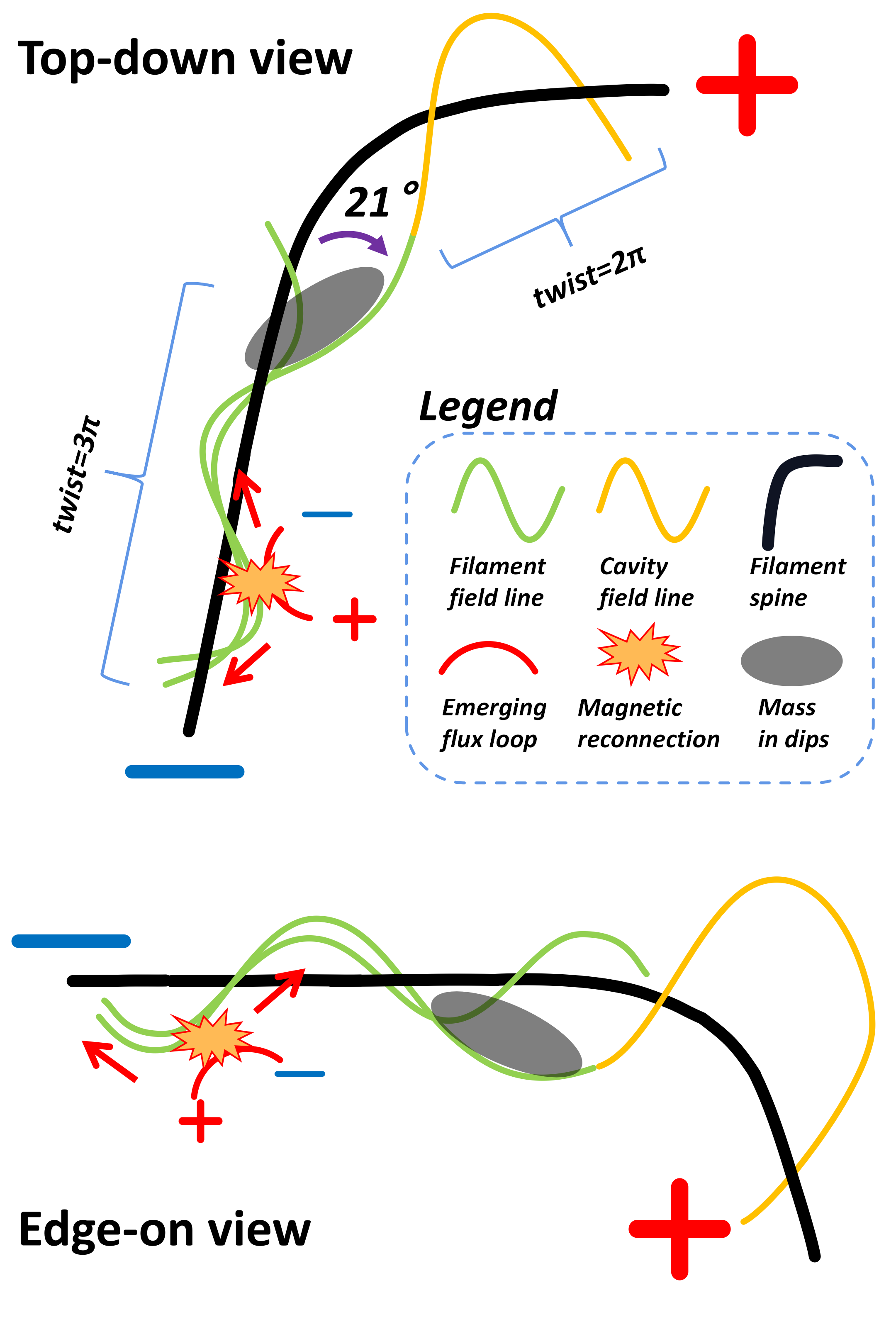}
\caption{Carton explaining inferred physics picture of jet-filament interaction. We plotted the cartoon in both directions to match the phenomena we observed from SDO and STEREO. The twist number of the filament and cavity is the conclusion of the \citetalias{10.1093/mnrasl/slac069}, noting the up-and-down relationship between the magnetic field lines and the filament spine. Further, based on the analysed direction of  filament oscillation, we hypothesize that an angle of 21 degrees exists between the magnetic lines of magnetic dips and the filament spine (indicated by the purple arrow).
}
\label{7}
\end{figure}

In particular, we use the cyan curve to represent the part of the void formed by the mass motion. For this part, we fit a circle to obtain the radius of curvature of this part (see Fig. \ref{6}(c)). We find a radius of curvature of 166 Mm at the formed void base, which is very close to the 151 Mm we obtain using the theoretical expression for longitudinal oscillations. This further validates the theory of filament LALOs and shows that the void we observed does correspond to the magnetic dip structures.

\subsubsection{Magnetic field strength}
\label{sec3.1.2}

The pendulum model of LALOs can also give us information about the minimum magnetic field strength of the dip. The minimum magnetic field strength of the dip can be obtained by bringing the period into the extended pendulum model of LALOs \citep{2022A&A...660A..54L}.
\begin{equation}
B \geq \frac{17P}{\sqrt{1-\left(\frac{P}{P_{\odot}}\right)^{2}}}
\end{equation}
The minimum magnetic field strength (23 G) is close to the previous report \citep{2014ApJ...785...79L,2014ApJ...790..100B,2017ApJ...850..143L,2017Ap&SS.362..165C}.

As the NF also exhibits LATO, we can likewise estimate the magnetic field strength in this part with the help of the oscillation parameters. According to the model of LATOs by \cite{1969SoPh....6...72K}, the strength of the effective magnetic field that causes the restoring force of the oscillation is given by 
\begin{equation}
B=4 \pi L P^{-1} \sqrt{\pi \rho} 
\end{equation}
, where  2L  is the length of the oscillating filament,  $\rho$  is the mass density, and  P  is the period  of the horizontal oscillation of a filament. Assuming  $\rho=10^{-13}\mathrm{~g} \mathrm{~cm}^{-1}$, B (21 G) can be calculated from the measured values of  $L=35 \mathrm{~Mm}  $ and  $P=0.33 \mathrm{~h} $. 

We can find that the magnetic field strength of the NF (21 G) is slightly less than that of the SF (23 G), which is quite reasonable since the NF is further away from the active region. This also indicates that the strength of the magnetic field inside a large filament is also heterogeneous.

\subsubsection{Jet energy}
\label{sec3.1.3}
TJ generated at the edge of the active area triggers mass oscillations in the filament we are studying, so we can get a rough estimate of the jet energy by estimating the kinetic energy of oscillations. The basic kinetic energy equation is $E=1 / 2 M_{\mathrm{f}} v^{2}$, where $M_{\mathrm{f}}$ is the mass of the oscillating filament and $v$ is the velocity amplitude of the oscillation. Considering that the oscillations are much more pronounced in the SF than in the NF, we consider only the longitudinal oscillations in the SF. From the H$\alpha$ image, we obtain that the length $L$ of the void is about 100 Mm and the width $W$ is 25 Mm. Assuming that the filament mass is stored in a homogeneous cylinder, the volume of the oscillating filament is $V=1 / 4 \pi W^{2} L$. Since the typical mass density of the filament is $\rho=10^{-13} \mathrm{~g} \mathrm{~cm}^{-1}$ and the total kinetic energy of the jet should be doubled (two-sided-loop jet), we can obtain
\begin{equation}
E=1 / 4 \rho \pi W^{2} L v^{2} = 1.96 \times 10^{28} \operatorname{erg}.
\end{equation}

Since we do not know what the conversion  rate of jet energy to oscillatory kinetic energy is, this energy estimate is only a lower bound on the jet energy. The jet energy estimated by filament seismology is in good agreement with the results of \cite{1999Ap&SS.264..129S} ($10^{25}-10^{29} \operatorname{erg}$) and \cite{2014ApJ...785...79L}($10^{24}-10^{27} \operatorname{erg}$). Because direct estimates of the energy of the TJ cannot be made, our seismological results provide a sample for the study of TJ.

\subsection{Magnetic structures of a filament-cavity system}
\label{sec3.2}

In the 2.5D MHD numerical experiment of \cite{2021ApJ...912...75L}, they set up a sheared magnetic arcade as the magnetic structure supporting the filament and studied the filament oscillations triggered by a coronal jet. This numerical simulation helps us understand our observational studies. In some filament magnetic field lines of their simulations, the jet impact was so strong that the filament mass in the magnetic dip is pushed out into the chromosphere. This is very similar to our observations, especially in Fig. \ref{1}(a3, b3 and c3). We can find that the mass in the SF deviated from the spine and even approached the northern part of the filament. Combined with stereoscopic observations from SDO and STEREO, we believe that the jet was pushing the mass in the filament magnetic dip into the coronal cavity to cause such a spectacular jet phenomenon. Because of the high height of the coronal cavity (about 150 Mm), some of the mass pushed out of the magnetic dip fell back into the magnetic dip, producing the LALOs we observe. A cartoon diagram of this has been made to complete our understanding of this physical process.

As shown in Fig. \ref{7}, the TJ originated near the filament's southern footpoint. The emerging flux loop in the filament channel was of opposite polarity to the filament above, and the reconnection produced bi-directional plasma flows (indicated by the red arrows). The jet, moving through the magnetic structure of the filament, exhibited a highly twisted structure, so we can diagnose a twist number of 3$\pi$ for this part filament. Most of the jet then deviated from the filament and pushed mass in the magnetic dip into the coronal cavity. A fraction of the mass that fell back undergoes oscillatory motion in the magnetic dip, and a fraction of the mass fell along the magnetic lines of the coronal cavity at the filament northern foot (this part magnetic structure of the cavity had a twist number of 2$\pi$).

In \citetalias{10.1093/mnrasl/slac069} we had the first insight into the magnetic rope structure of this filament-cavity system by tracing the path of TJ. Through the stereoscopic diagnosis of simultaneous longitudinal and transverse
oscillations, we get a more comprehensive picture of the magnetic structure. Further, we can see that the magnetic field lines of the coronal cavity are involved in the composition of the magnetic dip of the filament, which in turn helps us to understand why the filament and the coronal cavity are different parts of the same structure \citep{2006JGRA..11112103G}. At the beginning of the jet's movement along the magnetic structure, there is no significant difference between the coronal cavity and the filament (magnetic dip). In other words, the magnetic dip is  the bottom of the cavity, and under certain circumstances (e.g. away from the active region, weakening of the overlying field), the magnetic field of cavity expand into a semi-circular magnetic structure.

\section{conclusions}
In this paper, we report the first stereoscopic observation of simultaneous longitudinal and transverse oscillations in a single filament. The magnetic reconnection between the emerging flux loop and the overlying filament caused the TJ. The north arm of TJ also simultaneously caused the formation and recovery of filament void, suggesting the existence of periodic filament mass motions. With the help of stereoscopic observation, we can accurately judge the true trajectory of the jet that triggers the filament oscillation.

Using the time-space slit analysis of H$\alpha$ and EUV data, we find the oscillation of different parts of the filament and carefully analyze the oscillation mode and direction. We conclude that there are a LALO with an angle
of about 21 degrees to the spine in the southern filament and a LATO in the northern filament. For LALO, the initial amplitude, velocity, period, and damping timescale are 12.96 Mm, $19.21 \mathrm{~km} \mathrm{~s}^{-1}$, 1.18 h, and 2.04 hr, respectively. For LATO, the initial amplitude, velocity, period, and damping timescale are 2.99 Mm, $15.66 \mathrm{~km} \mathrm{~s}^{-1}$, 0.33 h, and 1.31 hr, respectively. The radius of curvature of the magnetic dip (151 Mm), the minimum magnetic field strength value (23 G) of the SF,  the effective magnetic field strength (21 G) of NF, and the jet energy ($9.81 \times 10^{27} \operatorname{erg}$) are obtained using the filament seismology analysis. We also fit a circle to the bottom of the
magnetic dip to obtain its radius of curvature (about 166 Mm). This qualitatively validates the theory of filament LALOs and shows that the void we observed does correspond to the magnetic dip structures.
Based on this, and in conjunction with the results of \citetalias{10.1093/mnrasl/slac069}, we flesh out the magnetic structures of filament. We conclude that emerging loops in filament channels often result in two-sided-loop coronal jets in the overlying filament owning to the magnetic reconnection between the two different types of magnetic structures. More importantly, the newly formed jet can not only trigger simultaneous longitudinal and transverse oscillations in a single filament, but also can be used as a seismology tool for diagnosing physical information of the filament, such as the magnetic structure, magnetic field strength, and magnetic twists.

A recent review of jet observations by \cite{2021RSPSA.47700217S} summarized the relationship between coronal jets and other solar eruptive phenomena. Coronal jets, which are generated near the footpoints of the filaments, are a good medium for triggering the motion of filament mass. There was also a growing number of studies of filament oscillations triggered by coronal jets \citep{2014ApJ...785...79L,2017ApJ...851...47Z,2021ApJ...912...75L}, which compels us to consider whether jet-filament interaction is a common mode for triggering filament oscillations? We will consider this question carefully in the following study.

\section*{Acknowledgements}

We thank the reviewer for careful reading of the manuscript and for constructive suggestions that improved the original version of the manuscript. We also would like to acknowledge the SDO, STEREO, and GONG science teams for providing the data. This work is supported by the Natural Science Foundation of China (11922307, 12173083, 11773068), the Yunnan Science Foundation for Distinguished Young Scholars (202101AV070004), the National Key R\&D Program of China (2019YFA0405000),  the Specialized Research Fund for State Key Laboratories, and the Yunnan Key Laboratory of Solar Physics and Space Science (202205AG07009).

\section*{Data availability}

This research used the SunPy \citep{sunpy_community2020}software package to present the results of our observation. All data is public and available through SunPy's download interface.
%%%%%%%%%%%%%%%%%%%%%%%%%%%%%%%%%%%%%%%%%%%%%%%%%%



%\bibliographystyle{mnras}
%\bibliography{ref} % if your bibtex file is called example.bib

%\begin{thebibliography}{99}
%\end{thebibliography}

%%%%%%%%%%%%%%%%%%%%%%%%%%%%%%%%%%%%%%%%%%%%%%%%%%

%%%%%%%%%%%%%%%%% APPENDICES %%%%%%%%%%%%%%%%%%%%%

%%%%%%%%%%%%%%%%%%%%%%%%%%%%%%%%%%%%%%%%%%%%%%%%%%

% Don't change these lines
\bsp	% typesetting comment
\label{lastpage}
\end{document}